\documentclass{jetpl}
\twocolumn
\title{Kopnin force and chiral anomaly}

\rtitle{Kopnin force and chiral anomaly}

\sodtitle{Kopnin force and chiral anomaly}

\author
{G.E. Volovik $^{+\#}$ \/\thanks{e-mail: volovik@boojum.hut.fi}
}

\rauthor{
G.E.Volovik
}

\sodauthor{
Volovik
}

\address{
$^{+}$ Olli Lounasmaa Laboratory, Aalto University, School of Science and
Technology, P.O. Box 15100, FI-00076 AALTO, Finland
\\
$^{\#}$ Landau Institute for Theoretical Physics RAS, Kosygina 2, 119334 Moscow, Russia
}

\abstract{ 
Kopnin spectral flow force acting on quantized vortices in superfluid and superconductors is discussed. Kopnin force represents the first realization of the chiral anomaly in condensed matter.
}

\begin{document}

\maketitle

 \section{Introduction}

\begin{figure}
\centerline{\includegraphics[width=1.0\linewidth]{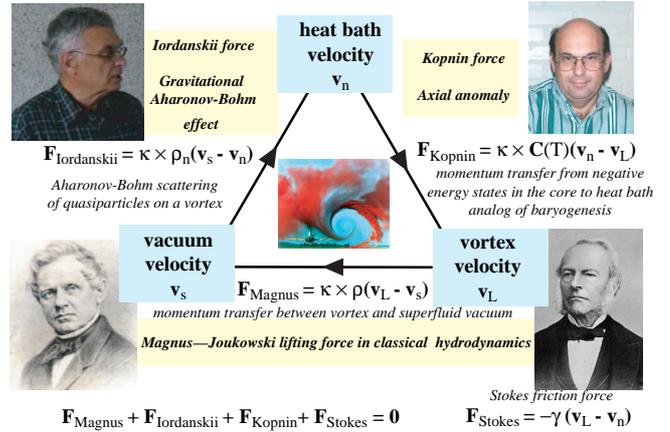}}
\caption{Fig. \ref{3Forces}. Balance of forces acting on a vortex line: three nondissipative forces and Stokes friction force. Here ${\bf \kappa}$ is the circulation quantum; $\rho$ is mass density of the liquid; $\rho_n$ and ${\bf v}_n$ are the normal component density and velocity; ${\bf v}_s$ is the superfluid velocity (velocity of ``superfluid vacuum''). The Kopnin force emerges in superconductors and fermionic superfluids. For vortex-skyrmion in chiral Weyl superfluids (Fig. \ref{Skyrmion}), the parameter $C$, which enters Kopnin force, is determined by Adler-Bell-Jackiw equation for chiral anomaly. In  $^3$He-A, which belongs to this class of topological materials,  the Kopnin parameter $C\approx \rho$. This prediction by Kopnin \cite{Kopnin1993} is confirmed in Manchester experiments \cite{Bevan1997}. For singular vortices in nonchiral superfluid $^3$He-B, the Kopnin parameter $C(T)$ depends on temperature $T$  \cite{Bevan1997}. It  changes from 0 at low $T$ to $\approx \rho$ at large $T>0.6 ~T_c$. In the latter limit, $C$ is fully determined by spectral flow and chiral anomaly. The reason for that is that the matter inside the core of the vortex in non-chiral liquid represents the chiral liquid with Weyl points (Fig. \ref{core}).  
}  
\label{3Forces} 
\end{figure}

A vortex moving in classical liquids experiences the famous Magnus or Kutta-Joukowski lift force. The vortex lines in a superfluid were introduced in seminal papers by L. Onsager and R. Feynman, and Abrikosov vortex lines in superconductors by A. Abrikosov. Quantized vortices in superfluids and superconductors experience two extra forces, which were introduced by Iordanskii and by  Kopnin:
Kopnin force \cite{Kopnin1976,Kopnin2001,Kopnin2002} and Iordanskii force 
\cite{Iordanskii1964,Iordanskii1966} (see Fig. \ref{3Forces}).

The Iordanskii force is a counterpart of the Magnus force produced by the normal component in the velocity field of a vortex moving with respect to the normal excitations. It exists both in Bose and Fermi superfluids and has an universal form which does not depend on any specific model: its value is fundamental being determined by the vortex winding number, density of the normal component and the vortex velocity relative to the normal component. The reason for universality is that the Iordanskii force has a topological nature and has the same origin as the gravitational Aharonov--Bohm effect in the presence of a spinning cosmic string \cite{Stone2000}.

The Kopnin force appears in the fermionic BCS systems: superconductors and fermionic superfluids. Vortices there confine low-energy bound states -- fermion zero modes. When the vortex moves, spectral flow of the excitation states is generated that transfers the vortex momentum into the environment and leads to a new type of transverse force on the vortex. The Kopnin spectral flow force is fundamental: it has the same origin as the chiral (axial) anomaly in relativistic quantum field theories. In chiral Weyl superfluid $^3$He-A, the Kopnin force acting on vortex-skyrmions is described by the Adler-Bell-Jackiw equation for chiral anomaly
\cite{AdlerBellJackiw}. The Kopnin force, representing the momentum transfer between the vortex and the normal component of the superfluid liquid, has full analogy with the transfer of baryonic charge in the phenomenon of electroweak baryogenesis \cite{Bevan1997}.

 %%%%%%%%%%%%%%%%%%%%%%%%%%%%%%%%%%%%%%%%
 %%%%%%%%%%%%%%%%%%%%%%%%%%%%%%%%%%%%%%%%
 %%%%%%%%%%%%%%%%%%%%%%%%%%%%%%%%%%%%%%%%
 %%%%%%%%%%%%%%%%%%%%%%%%%%%%%%%%%%%%%%%%

In modern cosmology, the phenomenon of baryogenesis is responsible for the excess of matter over anti-matter in our Universe. So, both the Iordanskii and Kopnin forces have a close relation to fundamental physics of elementary particles and gravity.
In a superconductor, the motion of a quantized Abrikosov vortex is the only significant source of dissipation. This makes the Iordanskii and Kopnin forces of considerable technological importance.
In superfluids, the temperature dependence of the transverse and friction forces calculated by Kopnin and confirmed by measurements in superfluid $^3$He-B  
\cite{Bevan1997} has led to the discovery of a unique phenomenon related to quantum turbulence of vortex lines. The transition to quantum turbulence  is governed by a new Reynolds number, the superfluid Reynolds number, that does not depend on velocity, but is determined by the ratio of transverse and drag forces 
\cite{Finne2003,Kopnin2004}.

While developed for condensed-matter physics, in neutron star physics the Iordanskii and Kopnin forces have greatly clarified and extended the understanding of vortex dynamics in neutron and proton superfluids. The phenomenon of spectral flow, which has been observed in superfluid $^3$He 
\cite{Bevan1997}, controls the drift velocity of vortices in the proton superconductor in a neutron star and so determines the rate at which magnetic flux can be expelled from the core to the crust \cite{Jones2009}. In quark matter, these two forces have dominant contributions to the dynamics of color-superconducting quark matter, which in particular may exist in the core of a neutron star \cite{Alford2010}.

In  Manchester experiments \cite{Bevan1997}  the calculated by Kopnin temperature dependence of the vortex forces have been measured both on singular vortices in $^3$He-B and on continuous vortices in chiral superfluid $^3$He-A. The latter  vortex structures are called skyrmions, which originally denoted the topologically twisted continuous field configurations in quantum field theory  \cite{Skyrme1962}. The quantized circulation around the continuous vortex is related to the topological winding number of the skyrmion texture in the field of the unit vector
$\hat{\bf l}$ along the direction of orbital angular momentum  \cite{Mermin-Ho}. 

\begin{figure}
\centerline{\includegraphics[width=0.8\linewidth]{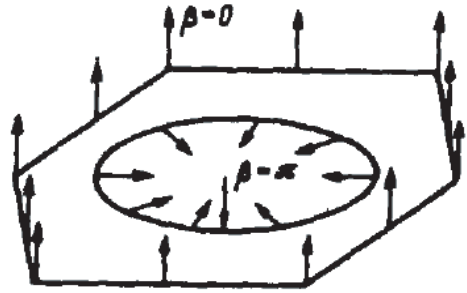}}
\caption{Fig. \ref{Skyrmion}. Simplified illustration of the elementary cell of the vortex  lattice from Ref. \cite{VolovikKopnin1977}. Each cell represents the vortex skyrmion, in which the unit vector $\hat{\bf l}$ along the direction of orbital momentum sweeps the $4\pi$ area on the unit sphere. Here $\beta$ is the polar angle of  vector $\hat{\bf l}$. 
}  
\label{Skyrmion} 
\end{figure}

Skyrmion lattices have been  first discussed in the Kopnin paper \cite{VolovikKopnin1977}, 
see Fig. \ref{Skyrmion}, where it has been suggested that skyrmions can be observed in $^3$He-A by NMR techniquie. And indeed, vortex-skyrmions have been experimentally  identified in $^3$He-A in NMR experiments \cite{Seppala1984}.
The lattice of skyrmions has been later discovered also in magnetic materials \cite{Muhlbauer2009}. In both $^3$He-A and in chiral magnetic systems an effective electrodynamics emerges, although from different origins, but in both cases this leads to an extra force acting on the skyrmions. In $^3$He-A, the configurations of skyrmions in the form of the vortex sheets have been also observed. The structure of the vortex sheet depends on preparation and is determined by dynamics of skyrmions along the sheet \cite{Kopnin2002}.

%%%%%%%%%%%%%%%%%%%%%%%%%%%%%%%%%%%%%%%%
 %%%%%%%%%%%%%%%%%%%%%%%%%%%%%%%%%%%%%%%%
 %%%%%%%%%%%%%%%%%%%%%%%%%%%%%%%%%%%%%%%%
 %%%%%%%%%%%%%%%%%%%%%%%%%%%%%%%%%%%%%%%%

Extending the theory of the Kopnin force to chiral superfluids with Weyl points, Kopnin already in 1991
predicted existence of fermionic bound states, which have exactly zero energy  \cite{Kopnin1991}. In our days such fermions are known as Majorana fermions - objects, which are still elusive in particle physics, but may be observable in topological superfluids and superconductors. Actually the paper   \cite{Kopnin1991} is in the origin of the consideration of Majorana fermions in condensed matter  \cite{ReadGreen2000}.

The zero energy bound states on vortices in chiral Weyl superfluids found by Kopnin have in addition the remarkable property: their spectrum is dispersionless. Now such flat bands are under intensive search in solid-state materials. According to recent works by Kopnin   
\cite{Kopnin2011,Kopnin2013,Kopnin2012}), the singular density of states in the materials with flat band may open the route to superconductivity at room temperature.

 \section{Kopnin force on vortex-skyrmion and chiral anomaly}

%%%%%%%%%%%%%%%%%%%%%%%%%%%%%%%%%%%%%%%%
 %%%%%%%%%%%%%%%%%%%%%%%%%%%%%%%%%%%%%%%%
\begin{figure}
\centerline{\includegraphics[width=0.6\linewidth]{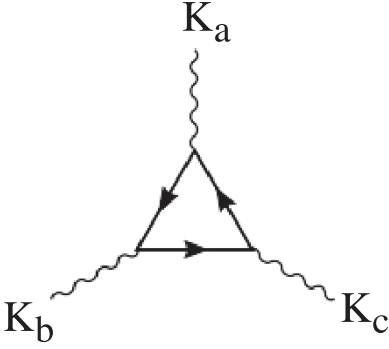}}
\caption{Fig. \ref{triangle}. Chiral anomaly in the 3+1 spacetime arises from triangle Feynman diagrams.
Here ${\rm K}_a$, ${\rm K}_b$  ${\rm K}_c$  correspond to the charges of three possible gauge fields 
acting on chiral fermions. In case of baryogenesis by hypermagnetic field  ${\rm K}_a=B$, the baryonic charge; and ${\rm K}_b={\rm K}_c=Y$, the hypercharge. In case of Kopnin force acting on continuous vortex skyrmion ${\rm K}_a={\bf p}^{(a)}$, the momentum of the Weyl point; ${\rm K}_b={\rm K}_c=Q$, the effective electric charge of fermions living in the vicinity of the Weyl points, $Q=\pm 1$.
}  
\label{triangle} 
\end{figure}
 %%%%%%%%%%%%%%%%%%%%%%%%%%%%%%%%%%%%%%%%
 %%%%%%%%%%%%%%%%%%%%%%%%%%%%%%%%%%%%%%%%

The chiral anomaly is the property of chiral fermions in the presence of external gauge and gravitational fields. In condensed matter, chiral fermions are studied in superfluid $^3$He-A, which has topologically protected Weyl points in quasiparticle spectrum. Chiral fermions also exist in the chiral matter, which lives in the core of quantized vortices. The chirality of the core matter is the consequence of the violated time reversal symmetry by the currents circulating around the core.
To illustrate the chiral anomaly origin of the Kopnin force, let us start 
with force acting on the vortex skyrmion in chiral Weyl superfluid, where the effect of chiral anomaly is more pronounced. 

The Weyl superfluid has Weyl points in the spectrum. Close to the Weyl points, quasiparticles behave as relativistic massless chiral fermions interacting with effective gravitational and gauge fields \cite{Volovik2003}. Since the chiral anomaly originates from the spectral flow through the Weyl points, and near the Weyl points the quasiparticles are relativistic, they experience the gauge and gravitational anomalies exactly in the same manner as elementary particles in relativistic quantum field theories. The gauge anomaly is described by  one-loop triangle Feynman diagram in Fig. \ref{triangle}, in which fermion is interacting with three gauge fields. These fields can be equal or different, so in general the fermions can be characterized by charges  $K_a,K_b,K_c$.
The charges are conserved, which means that the matrices of charges commute with the Green's function. That is why one can introduce the symmetry protected topological invariant, which contains the matrices $K_a,K_b,K_c$:
\begin{eqnarray}
N_{{\rm K}_a{\rm K}_b{\rm K}_c} = \frac{e_{\alpha\beta\mu\nu}}{24\pi^2} \times
\nonumber
\\
{\bf tr}\left[{\rm K}_a {\rm K}_b {\rm K}_c\int_\sigma   dS^\alpha
 G\partial_{p_\beta} G^{-1}
G\partial_{p_\mu} G^{-1} G\partial_{p_\nu}  G^{-1}\right].
\label{MasslessTopInvariantStandard Model}
\end{eqnarray}
Here $\sigma$ is the 3D surface around  the Weyl points in the 4D  frequency-momentum space
$p_\mu=(\omega,{\bf p})$; and ${\bf tr}$ denotes the trace over all the indices of the matrix Green's function $G(p_\mu)$. 

Using this topological invariant, one can write any type of the Adler-Bell-Jackiw equation for chiral anomaly. For example, the production
rate of baryonic charge in the presence of hyperelectric field ${\bf E}_{Y}$ and
hypermagnetic field ${\bf B}_{Y}$  is 
\begin{equation}
\dot B({\rm Y})=\frac{1}{4\pi^2}N_{{\rm B} {\rm Y} {\rm Y}} ~{\bf B}_Y\cdot {\bf E}_{Y} \,.
\label{BarProdByHypercharge}
\end{equation}
The baryoproduction is determined  the topological invariant $N_{{\rm B} {\rm Y} {\rm Y}} $ in Eq.(\ref{MasslessTopInvariantStandard Model}), where the charge
${\rm K}_a={\rm B}$ is the baryonic charge; and charges ${\rm K}_b$ and ${\rm K}_c$ are equal to the hypercharge ${\rm Y}$.

The similar equation is applicable for the production of another fermionic charge -- the quasiparticle linear momentum ${\bf p}$.
In Weyl superfluids each Weyl point has its momentum -- the position 
${\bf p}^{(a)}$  in momentum space. The production rate of the momentum is  the force. In case of a moving vortex this is the Kopnin force. 

In $^3$He-A, the positions of the Weyl points is determined by the orbital momentum vector, ${\bf p}^{(a)}=\pm p_F\hat{\bf l}$.   Substituting into Eq.(\ref{BarProdByHypercharge}) the relevant fermionic charges  (${\rm K}_a={\bf p}^{(a)}$, the momentum of the Weyl point, and ${\rm K}_b={\rm K}_c=Q$, the effective electric charge of fermions living in the vicinity of the Weyl points), one
obtains that the rate of momentum production induced by the time and space dependent $\hat{\bf l}$-texture is:
\begin{equation}
\dot{\bf P} = \frac{1}{4\pi^2}{\bf N}_{{\bf p}^{(a)} {\rm Q} {\rm Q}}  \left({\bf B} \cdot {\bf E} \right) \,.
\label{MomentumProductionGeneral}
\end{equation}
Here ${\bf B}=(p_F/\hbar)\nabla\times\hat{\bf l}$ and   ${\bf E}= (p_F/\hbar)
\partial_t
\hat{\bf l}$ are effective `magnetic'  and `electric' fields acting on
Weyl quasiparticles in $^3$He-A;
${\rm Q}=\pm 1$ is the matrix of corresponding `electric' charges. In $^3$He-A the `electric' charge is opposite to the chirality of
the Weyl quasiparticle.
As a result,  the
momentum production by the $\hat{\bf l}$-texture per unit time per unit volume is
\begin{equation}
 {\dot{\bf P}}=- \frac{p_F^3}{2\pi^2 \hbar^2}\hat{\bf l}
\left(\partial_t\hat{\bf l}\cdot(\nabla\times \hat{\bf l})\right)
   \,.
\label{MomentumProductionAPhase}
\end{equation}
Integrating this equation (\ref{MomentumProductionAPhase}) over the cross-section of the vortex-skyrmion in Fig. \ref{Skyrmion} moving with velocity 
${\bf v}_{\rm L}$  with respect to the heat bath, one obtains the Kopnin force acting on skyrmion:
\begin{equation}
F^{\rm Kopnin}_i = - 2\pi\hbar~ \frac{p_F^3}{3\pi^2 \hbar^3} {\hat {\bf z}}  \times {\bf v}_{\rm L} \,.
\label{KopninForce}
\end{equation}

Let us compare the Kopnin force with the traditional Magnus force determined by the particle density $n$ (see Fig. \ref{3Forces}, where the mass densitty $\rho=mn$):
\begin{equation}
{\bf F}^{\rm Magnus}=   2\pi \hbar ~ n ~{\hat {\bf z}}  \times {\bf v}_{\rm L} \,.
\label{MagnusForce}
\end{equation}
 In the weak coupling BCS regime, the density of the liquid $n$ in its superfluid state only slightly deviates from the density of the liquid  in the normal state, $n_{\rm normal~state} = p_F^3 / 3\pi^2 \hbar^3$. That is why from Eqs.  (\ref{KopninForce}) and (\ref{MagnusForce}) it follows  that in $^3$He-A, where the weak coupling regime is applicable, the Kopnin force should practically compensate the Magnus force \cite{Kopnin1993}. This compensation has been observed in Manchester experiment on $^3$He-A \cite{Bevan1997}. 

 \section{Kopnin force on singular vortex}

 For the singular vortices the Kopnin force is also connected with spectral flow. But as distinct from the case of continuous vortex-skyrmions, the spectrum of 
 fermionic quasiparticles in the vortex core has minigap $\omega_0(p_z)$. Due to discrete spectrum the spectral flow is suppressed. As a result the parameter $C$, which enters the Kopnin force in Fig. \ref{3Forces}, is smaller than the mass density 
 $\rho$ and depends on temperature.  The Kopnin function $C(T)$ in Fig. \ref{3Forces} approaches zero in the limit of zero temperature, and approaches the spectral flow value $\approx \rho$ at $T>0.6~ T_c$  \cite{Bevan1997}. The latter happens because at high temperature  the distance between the core levels is smaller than the width of the level due to dissipation, $\omega_0\tau \ll 1$, and  the spectral flow in the vortex core  is restored. 
 
 %%%%%%%%%%%%%%%%%%%%%%%%%%%%%%%%%%%%%%%%
 %%%%%%%%%%%%%%%%%%%%%%%%%%%%%%%%%%%%%%%%
\begin{figure}
\centerline{\includegraphics[width=0.9\linewidth]{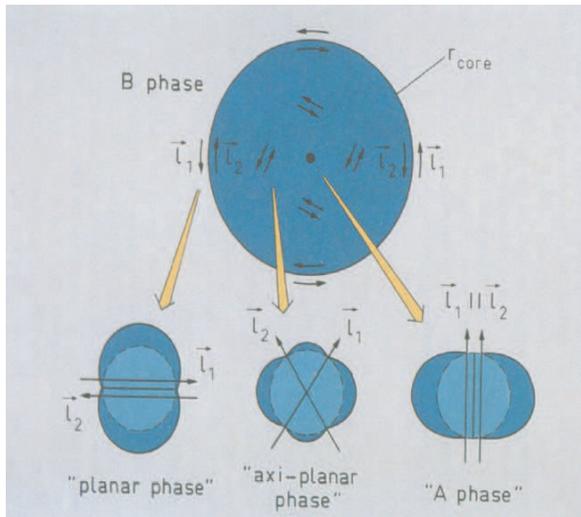}}
\caption{Fig. \ref{core}.  Illustration of the core structure of axisymmetric vortex in $^3$He-B (from Ref. \cite{SalomaaVolovik1987}).  The core matter represents the chiral superfluid with 4 Weyl points
in positions 
${\bf p}^{(a)}=(\pm p_F\hat{\bf l}_1,\pm p_F\hat{\bf l}_2)$. The chiral anomaly, which comes from the chiral fermions living near the Weyl points, is responsible for the Kopnin force acting on a moving vortex.
}  
\label{core} 
\end{figure}
 %%%%%%%%%%%%%%%%%%%%%%%%%%%%%%%%%%%%%%%%
 %%%%%%%%%%%%%%%%%%%%%%%%%%%%%%%%%%%%%%%%

The connection of the Kopnin force with the chiral anomaly equations in the limit  $\omega_0\tau \ll 1$ can be visualized in the semiclassical limit, when momentum ${\bf p}$ and coordinate ${\bf r}$ can be considered as independent variables of Green's function. Such consideration is applicable if the core size exceeds the coherence length \cite{Kopnin1993}. The structure of the core of a singular vortex in $^3$He-B  is illustrated in Fig. \ref{core}.  Inside the core, the quasiparticle spectrum in semiclassical approximation contains four Weyl points (two per each spin component) in positions 
${\bf p}^{(a)}=(\pm p_F\hat{\bf l}_1,\pm p_F\hat{\bf l}_2)$.
Since the core matter represents the chiral superfluid with Weyl points, the 
Eq.(\ref{MomentumProductionGeneral}) is applicable. Then integrating Eq.(\ref{MomentumProductionAPhase}) over the cross-section of the vortex core, one obtains the Kopnin force in Eq.(\ref{KopninForce}) (with extra  factor $1/2$ if one considers the singly-quantized vortex instead of the doubly quantized vortex-skyrnion in $^3$He-A). 
 
  \section{Conclusion}

Kopnin force represents the first realization of the chiral anomaly in condensed matter.

Similar gauge anomalies are now popular in the other systems, such as the dense quark-gluon matter in QCD and hypothetical Weyl materials -- Weyl semimetals, see 
Refs. \cite{Basar2013,BasarYee2013,Son2012,Son2013,VivekAji2012}.
 
\section*{\hspace*{-4.5mm}ACKNOWLEDGMENTS}
I acknowledge financial
support  by the EU 7th Framework Programme
(FP7/2007-2013, grant $\#$228464 Microkelvin) and by the Academy of
Finland through its LTQ CoE grant (project $\#$250280).

 \end{document}